\documentstyle [12pt]{article}

\parskip=\medskipamount                 
\def\7#1#2{\mathop{\null#2}\limits^{#1}}        

\def\beee{\begin{equation}}
\def\eeee{\end{equation}}

\begin{document}

\bibliographystyle{unsrt}

\begin{center}

{\Large \bf COVARIANT SINGLE-TIME\\ BOUND-STATE EQUATION}
\footnote{Supported in part by the National Science Foundation.\\
e-mail addresses, greenberg@umdep1.umd.edu, rashmi@greta.ecm.ub.es,
schlumpf@wam.umd.edu\\
Present address of R. Ray: Facultat de Fisica, Universitat
de Barcelona, 08028 Barcelona, Spain.}\\
[5mm]
O.W. Greenberg, R. Ray and F. Schlumpf\\
{\it Center for Theoretical Physics\\
Department of Physics\\
University of Maryland\\
College Park, MD~~20742-4111}\\[5mm]

Preprint number 95-120\\

hep-ph/9504396\\

\end{center}

\vspace{2mm}

\begin{center}
{\bf Abstract}
\end{center}

We derive a system of covariant single-time equations for a two-body bound
state in a
model of scalar fields $\phi_1$ and $\phi_2$ interacting via exchange of
another
scalar field $\chi$.  The derivation of the system of equations follows from
the Haag expansion.  The equations are linear integral equations that are
explicitly symmetric in the masses, $m_1$ and $m_2$, of the scalar
fields, $\phi_1$ and $\phi_2$.  We present an approximate analytic formula for
the mass eigenvalue of the ground state and give numerical results for the
amplitudes for a choice of constituent and exchanged particle masses.

\newpage
{\bf 1. INTRODUCTION}

The problem of relativistic bound states has a long history.  Nonetheless, the
treatment of this problem is still not
completely satisfactory.  The purpose of this
paper is to continue the development of an alternative to the most popular
formulation, the
Bethe-Salpeter method\cite{bs}. The Bethe-Salpeter method
uses amplitudes in which both constituents are
off-shell.  Because of this, the amplitudes depend on an unphysical
relative-time coordinate and obey equations that are difficult to solve and
have spurious unphysical solutions, including some of negative norm.
Several authors have proposed covariant,
single-time equations with only one constituent off-shell.  The equations most
similar to ours are the ``spectator
equations'' of F. Gross\cite{fg}.  Our equations differ from the spectator
equations in the way we ensure symmetry between the off-shell and on-shell
particles, in the inclusion of renormalization graphs and counter terms and in
the boundary conditions of the Green's functions: our equations use Green's
functions with retarded boundary conditions, rather than Feynman boundary
conditions.  Our derivation of the equations differs entirely from Gross'
derivation of the spectator equation:  we use the Haag expansion\cite{rh}
and the operator field equations\cite{owg,gg,cg,gm,orr},
rather than summing classes of Feynman graphs.  To be concrete, we consider a
two-body bound state in a model of scalar fields $\phi_1$ and $\phi_2$
interacting via exchange of another scalar $\chi$.  A.
Raychaudhuri used this method to study bound states in the equal-mass case of
this model ; however his equations are not symmetric
in the on-shell and off-shell masses\cite{r1}.
This asymmetry was not evident in the equal-mass case.  He also studied the
nonrelativistic
reduction of the equations for unequal-mass bound states of spin-$1/2$
particles\cite{r2}.  Related work was done by M. Bander, et al\cite{ba}.

In this paper, we extend Raychaudhuri's analysis to unequal-mass constituents
and treat the on-shell and off-shell particles in a completely symmetric
way.  We present numerical results for the ground
state eigenvalue and amplitudes for a range of
$m_1/m_2$ and $\mu/m_2$, where $\mu$ is the mass of the $\chi$ field.

We hope this method will provide an alternative to the Bethe-Salpeter method
with the following advantages: (1) because only one particle at a time is
off-shell, the amplitudes depend on only one invariant, (2) all normalizable
solutions are physical and have positive norm, (3) the limit for one mass very
large is the relativistic equation for the other particle bound in an external
field and (4) the nonrelativistic limit has the correct reduced mass.
We have two longer-range goals: (1) to extend the Haag expansion to account
for cases in which the interaction is strong enough to make a two-particle
description inadequate and (2) to modify the Haag expansion to treat confined
degrees of freedom, for which the usual asymptotic fields don't exist.  This
latter goal will require a significant generalization of the method.

{\bf 2. DERIVATION OF THE EQUATIONS}



In this section we obtain coupled integral equations for two bound-state
amplitudes, one, $f_1$, with particle one off-shell and particle two on-shell,
the other, $f_2$, with these roles interchanged.   Our Lagrangian is
\begin{equation}
{\cal L}=\sum_{i=1}^2\frac{1}{2}(\partial_\mu \phi_i \partial^\mu \phi_i) -
m_i^2 \phi_i^2)
 + \frac{1}{2}(\partial_\mu \chi \partial^\mu \chi - \mu^2 \chi^2)
 + \frac{g}{4}[\phi_1^2+\phi_2^2,\chi ]_+ ,
\end{equation}
where the last term is an anticommutator.
We work in momentum space using $\phi(x)=(2 \pi)^{-3/2}\int d^4p
\tilde{\phi}(p) exp(-i p \cdot x)$ and the analogous formula with in fields.
We
promptly drop the tilde on $\phi$, abbreviate $d^4p$ by $dp$ and, for the in
field, abbreviate $:\phi^{in}(p) \delta(p^2-m_i^2):$ by $:\phi^{in}(p):$.
The equations of motion in momentum space are
\begin{eqnarray}
	(m_i^2-p^2)\phi_i(p) & = & \frac{g}{2(2 \pi)^{3/2}}
\int dp_1 dp_2 \delta(p-p_1-p_2)
	[\phi_i(p_1),\chi(p_2)]_+ \\
& &  +(A_i p^2 -B_i m_i^2)\phi_i(p)\\
	(\mu^2-p^2)\chi(p) & = & \frac{g}{2(2 \pi)^{3/2}}
\sum_{i=1}^2 \int dp_1 dp_2
	\delta(p-p_1-p_2) \phi_i(p_1) \phi_i(p_2) \\
& &+(D p^2 -E \mu^2)\chi(p),
\end{eqnarray}
where we have introduced counter terms for the mass and field strength
renormalizations of the fields.

In the $N$-quantum approximation, we expand the Lagrangian
fields in terms of the complete, irreducible set of in-fields (or out-fields),
including those for stable bound states (this is the Haag expansion),
and truncate the expansion to find an approximate set of equations among a
finite number of amplitudes.  Here we keep all terms that contribute to
equations for the two-body bound-state amplitudes $f_1$ and $f_2$ mentioned
above in one-loop approximation.  All the terms have explicit order $g^2$ at
the
perturbative vertices.  The relevant terms in the Haag expansion are

\newpage

\begin{eqnarray*}
	\phi_1(p) & = & :\phi_1^{in}(p): +
	\int dq db \delta(p+q-b)f_1(q,b):\phi_2^{in}(-q) B^{in}(b):  \\
	 &  & +\int dq dl db \delta(b-q-l-b)f_{\chi 2B}(q,b,l)
	 :\chi^{in}(q)\phi_2^{in}(l)B^{in}(b):,
\end{eqnarray*}
\begin{eqnarray*}
	\chi(p) & = & :\chi^{in}(p): + \int dl_1 dl_2
	\delta(p-l_1-l_2)\gamma_{ii}(l_1,l_2):\phi_i^{in}(l_1)
	\phi_i^{in}(l_2)  \\
	 &  & + \int dl_1 dl_2 db \delta(p-l_1-l_2-b)\gamma_{12B}(l_1,l_2,b)
	 :\phi_1^{in}(l_1)\phi_2^{in}(l_2)
	 B^{in}(b): ,
\end{eqnarray*}
and the analogous terms for $\phi_2$, with 1 and 2 interchanged.  Our general
notation is that the subscript on an amplitude lists its associated product of
in fields.  What we call $f_1$ should be $f_{2B}$ according to this general
notation; however, for convenience, we call it $f_1$. Because each in field
has a mass shell $\delta$-function, all the momentum integrals in the Haag
expansion are on two-sheeted mass hyperboloids.  In $f_1$ we keep $b$ on the
positive-energy mass shell and reverse the sign
of the momentum $q$ in $f_1$ so that $q$ on the positive-energy hyperboloid
gives the dominant amplitude in the nonrelativistic limit.  We call $f_1(q,b)$
with $q>0$, i.e. with $q$ on the positive mass hyperboloid, $f_1^{(+)}$ and
with
$q<0$, $f_1^{(-)}$.  (See Fig. 1)  Both our equations and their graphical
representation include both of these pieces of the amplitudes; to save space we
don't exhibit both pieces in the graphs.

As usual, $:\: :$ denotes normal ordering. In the one-loop approximation,
contractions always involve the vacuum matrix element of the anticommutator,
\begin{eqnarray*}
\langle [\phi_i^{in}(p_1), \phi_j^{in}(p_2)]_+\rangle_0 & = &
\delta_{ij} \delta(p_1+p_2) \delta_{m_i}(p_1) ,  \\
\langle [\chi^{in}(p_1), \chi^{in}(p_2)]_+\rangle_0 & = &
\delta(p_1+p_2) \delta_{\mu}(p_1),  \\
\end{eqnarray*}
where $\delta_{m}(p)=\delta(p^2-m^2)$ for short.
Choosing to expand the Lagrangian fields in terms of the
in-fields requires using retarded boundary conditions for the propagators.

To obtain the equation for $f_1$, we insert the Haag expansions for $\phi_1$
and $\chi$ in the equation of motion for $\phi_1$, renormal order and equate
the
coefficients of the term with $:\phi_2^{in}B^{in}:$.  The resulting equation
involves the amplitudes $f_1,~ f_{\chi 2B},~ \gamma_{22},$ and $\gamma_{12B}$.
We calculate the last three amplitudes in terms of $f_1$ using the equations of
motion and the Born approximation for emission of both on-shell and off-shell
$\chi$ quanta.  We will give details of this in a later, more detailed paper.
\begin{eqnarray}
f_{\chi 2B }(q,l,b) & = & \frac{g}{(2 \pi)^{3/2}}\,
\frac{f_1(-l,b)}{m_1^2-(q+l+b)^2},\\
\gamma_{11}(p_1,p_2) & = & \gamma_{22}(p_1,p_2) =
\frac{g}{2(2 \pi)^{3/2}}\,
\frac{1}{\mu^2-(p_1+p_2)^2} ,
\\
\gamma_{12B}(l_1,l_2,b) & = & \frac{g}{(2 \pi)^{3/2}}\,
\frac{f_1(-l_2,b)+f_2(-l_1,b)}{\mu^2-(l_1+l_2+b)^2}.\\
\end{eqnarray}
The integral equation for $f_1$ is
\begin{eqnarray}
  \lefteqn{[m_1^2-(b-p)^2] f_1(p, b)=}\nonumber\\
  && \frac{g^2}{16 \pi^3} \int dq \left[
  \frac{ \delta_{m_1}(q)}{\mu^2-(b-p-q)^2}+
  \frac{ \delta_{\mu}(q)}{m_1^2-(b-p-q)^2} \right]f_1(q,b) \nonumber\\
  && + \frac{g^2}{16 \pi^3} \int dq \left[
  \frac{ \delta_{m_2}(q)}{\mu^2-(p-q)^2}f_1(q,b)+
  \frac{ \delta_{m_1}(q)}{\mu^2-(b-p-q)^2}f_2(q,b)\right] \nonumber\\
  && +[A_1(b-p)^2 -B_1m_1^2]f_1(p,b),
  \label{eq:1}
\end{eqnarray}
where the first two terms on the right hand side are self-energy graphs
that are completely canceled by the renormalization counter terms.  Any method
of regularization of the self-energy graphs will suffice.  The third
and fourth terms on the right give binding by exchange of the $\chi$ field.
The bound-state momentum $b$ is always on its mass shell $b^2=M^2$.
This bound-state equation is shown in Fig. 2.
We get another coupled equation for $f_2$ by
interchanging $1$ and $2$.  The resulting pair of equations is clearly
symmetric
under $1$, $2$ interchange.
We suppress the $i\epsilon$'s associated with the retarded boundary conditions.

{\bf 3. Approximate Mass Eigenvalue Formula}

We considered parametrizing the mass eigenvalue formula using the
$\arccos \eta$,
where $\eta=M/(m_1+m_2)$, because this expression appears in the hyperboloidal
harmonic analysis that Raychaudhuri\cite{r1} used.
We found interesting empirical
regularities using this parametrization. The result is
\begin{equation}
M=(m_1+m_2) \cos \frac{\lambda - a}{b},
\end{equation}
where $\lambda=g^2/(32 \pi m_1m_2)$, $a=0.9 \sqrt{\mu/m_{red}}$ and
$b= 0.8-1.1~ \ln (m_</m_>)$.  The reduced mass is the usual expression;
$m_>$ is the larger of $m_1$ and $m_2$.  The range of validity of this
empirical formula is
$0 \leq \mu \leq m_<$, $0.01 \leq m_</m_> \leq 1$, $0.5 \leq
\eta \leq 1$ for $m_</m_>=1$ and $0.9 \leq \eta \leq 1$ for $m_</m_>=0.1$.

{\bf 4. Numerical Results}

Equation~(9) and the one with $m_1$ and $m_2$ interchanged
are eigenvalue equations
for the coupling constant $g$; that means for given values of the masses
$M, m_1, m_2$ and $\mu$ we can find a coupling constant $g$ and
wave functions $f^{(\pm)}_{1,2}$ that satisfy the equations. We solve these
homogeneous linear integral equations by approximating
the integral on the right hand side
with a finite sum. We choose Gauss integration with
appropriate points and weights. The resulting matrix equation
is solved by standard means.

For the equation in momentum space it is sufficient to take 18 mesh
points to obtain $g^2$ to an accuracy of 4\%. The main
difficulties we encounter in Eq.~(9) are the
logarithmic singularities.  We smooth these singularities by
keeping a finite $\epsilon$ at the logarithmic singularity.
We checked that the result does not change by varying
the mesh points and $\epsilon$.

In Fig. 3 the value of $\lambda=g^2/(32\pi m_1 m_2)$ is
plotted as a function of $\eta=M/(m_1+m_2)$ for $m_</m_>=0.1$ and $\mu=0$.
A calculation using the Bethe -Salpeter equation~\cite{zlm69} consistently
gives smaller binding.  In the scalar model we cannot
decide which solution is correct because there is no
experimental data.
Figure~4 shows the wave functions $f_1^{(+)}$, $f_1^{(-)}$, $f_2^{(+)}$ and
$f_2^{(-)}$,
respectively, for the mass ratios given above and for $\eta=0.95$.
As expected $f_1^{(+)}$ is the dominant contribution.

We will present more extensive numerical results in a later paper.

{\bf 5. Summary and outlook for future work}

The Haag expansion leads directly to coupled linear
integral equations for four amplitudes related to a scalar bound state.  Two
amplitudes are those that reduce to the nonrelativistic
wavefunction, one, $f_1$, with the particle of mass $m_1$ off-shell
and the other, $f_2$, with
the particle of mass $m_2$ off-shell.  The other two amplitudes have the
on-shell
particle crossed, so that its momentum lies in the same light cone as the
bound-state momentum.  These four amplitudes obey a set of four coupled linear
integral equations.  We solved these numerically
using momentum-space variables.

We plan to apply this method to bound states of two spin-$1/2$ particles, such
as the hydrogen atom and positronium, where
our calculations can be compared with experimental results. We hope this method
can replace the Bethe-Salpeter method in theories without confinement.

In order to use this method in confining theories, such as QCD, the asymptotic
fields that are a prominent part of the Haag expansion must be replaced with
fields that correspond to confined degrees of freedom.  The treatment of
confined degrees of freedom in theories such as
QCD remains a goal for the future.

{\bf Acknowledgements}

It is a pleasure to thank Joe Sucher and Boris Ioffe
for illuminating discussions.

{\bf Captions}

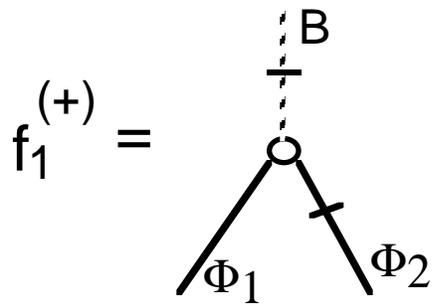
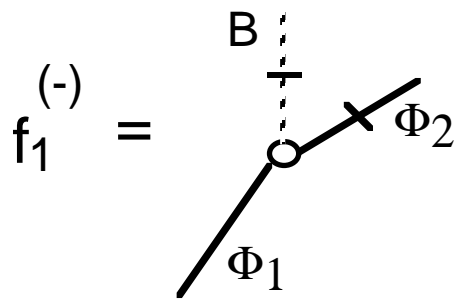
Fig. 1:  The two pieces of $f_1$.  The short line through a leg indicates the
leg is off-shell.

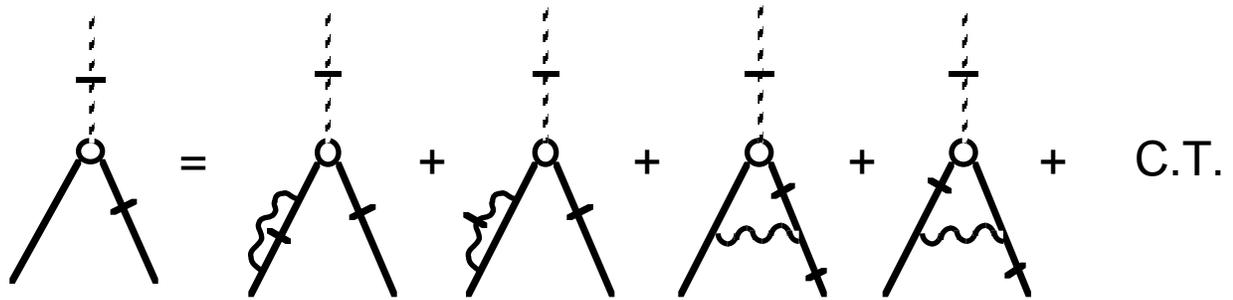
Fig. 2:  Graphs for the bound state equation for $f_1$ if the left-hand leg is
$\phi_1$.  Note that the first two terms are self-energy graphs that are
canceled by the counter terms, the third term is a $t$-channel graph that
couples $f_1$ to itself and the last term is a $u$-channel graph that couples
$f_1$ to $f_2$.

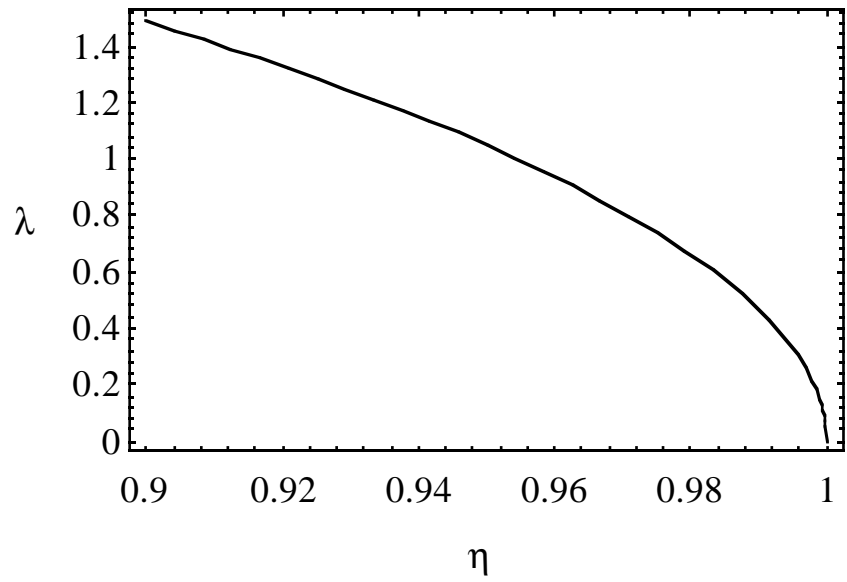
Fig. 3:  Plot of $\lambda=g^2/(32\pi m_1 m_2)$ as a function of
$\eta=M/(m_1+m_2)$ for $\mu=0,~m_</m_>=0.1$.

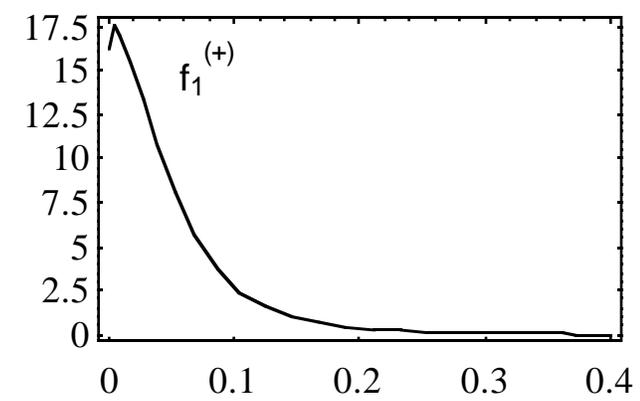
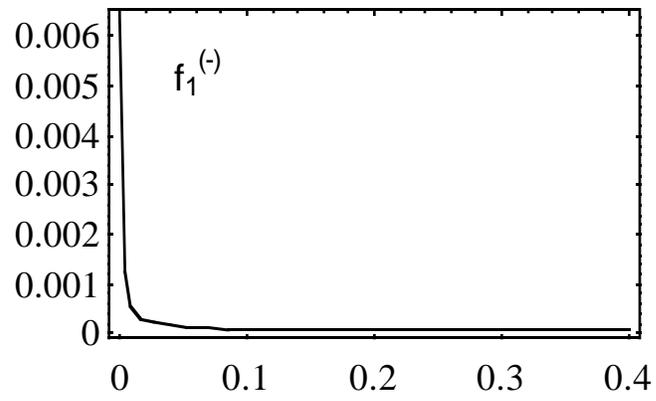
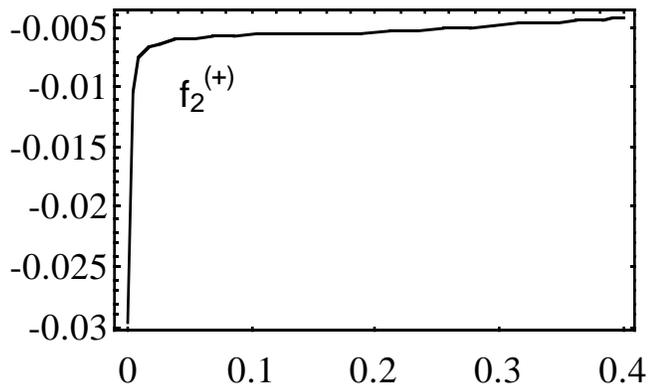
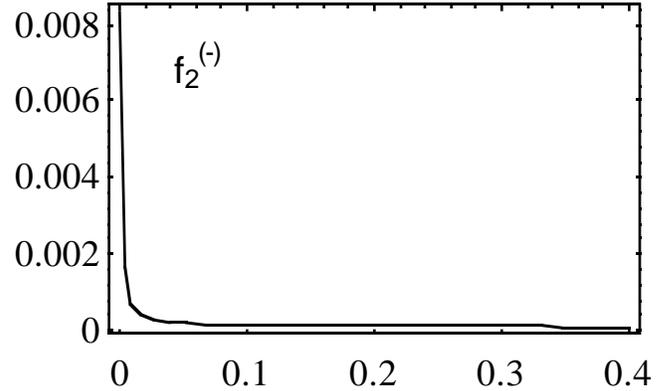
Fig. 4:  Wave functions $f_1^{(+)}$, $f_1^{(-)}$, $f_2^{(+)}$ and
$f_2^{(-)}$ in momentum space in arbitrary units as a function of $\Lambda$ for
$\eta=0.95$, $\mu=0$ and $m_</m_>=0.1$.
Here $m_>\cosh\Lambda=\sqrt{{\bf p}^2+m_>^2}$.

\end{document}